\documentclass[11pt]{article}

\usepackage{amsmath}

\usepackage{times}
\usepackage{bm}
\usepackage{natbib}

\usepackage[left=2.5cm,right=2.5cm,top=2.5cm, bottom=4cm]{geometry}
\usepackage[plain,noend]{algorithm2e}

\makeatletter
\renewcommand{\algocf@captiontext}[2]{#1\algocf@typo. \AlCapFnt{}#2} 
\def\@algocf@capt@plain{top}
\renewcommand{\algocf@makecaption}[2]{%
  \addtolength{\hsize}{\algomargin}%
  \sbox\@tempboxa{\algocf@captiontext{#1}{#2}}%
  \ifdim\wd\@tempboxa >\hsize
    \hskip .5\algomargin%
    \parbox[t]{\hsize}{\algocf@captiontext{#1}{#2}}
  \else%
    \global\@minipagefalse%
    \hbox to\hsize{\box\@tempboxa}
  \fi%
  \addtolength{\hsize}{-\algomargin}%
}

\makeatother

\def\P{{\text{pr}}}

\usepackage{caption}
\usepackage{subcaption}
\usepackage{tikz}
\usepackage{url}
\usepackage[thmmarks]{ntheorem}

\theoremheaderfont{\hskip\parindent\normalfont\scshape}
\theorembodyfont{\itshape}
\theoremseparator{.}

\theoremnumbering{arabic}
\newtheorem{theorem}{Theorem}
\theoremnumbering{arabic}

\theoremnumbering{arabic}

\theoremnumbering{arabic}

\theoremnumbering{arabic}

\theoremnumbering{arabic}
\theoremheaderfont{\hskip\parindent\normalfont\itshape}
\theorembodyfont{\rm}

\theoremnumbering{arabic}

\theoremnumbering{arabic}

\theoremnumbering{arabic}

\theoremnumbering{arabic}

\theoremnumbering{arabic}

\theoremheaderfont{\hskip\parindent\normalfont\itshape}
\theorembodyfont{\normalfont}

\theorembodyfont{\itshape}
\theoremheaderfont{\hskip\parindent\normalfont\scshape}
\newtheorem{result}{Result}

\font\QEDlogofont=msam10 at 10pt
\def\QEDlogo{\hbox{\QEDlogofont\char'003}}

\theoremsymbol{\QEDlogo}
\theoremheaderfont{\hskip\parindent\normalfont\itshape}
\theorembodyfont{\normalfont}
\theoremseparator{.}
\theoremstyle{nonumberplain}

\renewtheoremstyle{nonumberplain}%
  {\item[\theorem@headerfont\hskip\labelsep ##1\theorem@separator]}%
  {\item[\theorem@headerfont\hskip \labelsep ##1\ ##3\theorem@separator]}

\begin{document}



\markboth{EE Gabriel et~al.}{Ordinal Bounds}

\title{Sharp symbolic nonparametric bounds for measures of benefit in observational and imperfect randomized studies with ordinal outcomes}

\author{Erin E. Gabriel \\ Section of Biostatistics, Department of Public Health, University of Copenhagen, Denmark, \\ erin.gabriel@sund.ku.dk \\
Michael C. Sachs \\ Andreas Kryger Jensen}

\maketitle

\begin{abstract}
The probability of benefit is a valuable and important measure of treatment effect, which has advantages over the average treatment effect. Particularly for an ordinal outcome, it has a better interpretation and can make apparent different aspects of the treatment impact. Unfortunately, this measure, and variations of it, are not identifiable even in randomized trials with perfect compliance. There is, for this reason, a long literature on nonparametric bounds for unidentifiable measures of benefit. These have primarily focused on perfect randomized trial settings and one or two specific estimands. We expand these bounds to observational settings with unmeasured confounders and imperfect randomized trials for all three estimands considered in the literature: the probability of benefit, the probability of no harm, and the relative treatment effect. \\
noncompliance; probability of benefit; probability of no harm; relative treatment effect; symbolic nonparametric bounds
\end{abstract}

\section{Introduction}
The probability of benefit and its variations, the probability of no harm and the relative treatment effect, are useful estimands for determining the impact of a treatment. Unlike the average causal effect, a large benefit in one patient subgroup cannot overwhelm a small amount of harm or no benefit for the majority of patient subgroups in the population of interest. Although it is commonly misunderstood, the probability of benefit, even in a randomized trial setting, is not identifiable without strong, untestable assumptions. In particular, identification requires knowledge about the joint distribution of potential outcomes, something that is considered unreasonable by some researchers \citep{greenland2020causal}. Several papers have made this point and provided nonparametric causal bounds for the probability of benefit and variations of it in randomized settings for ordinal and continuous outcomes \citep{fan2010sharp, lu2018treatment, lu2020sharp, fay2018causal, huang2016inequality}. 

\citet{lu2018treatment} consider the probability of no harm, i.e., the probability that individuals have greater or equal outcomes under treatment than control (assuming that larger values of the outcome are better), and the probability of benefit, the probability that the outcome is strictly greater 
under treatment than control; we define these two estimands more precisely via potential outcomes in the next section. They derive bounds for these quantities that depend on the marginal probabilities of potential outcomes. The marginal quantities can be estimated in randomized trials, or in cases where the treatment assignment is determined from observed covariates, i.e. under the assumption of no unmeasured confounders. They also consider covariate-adjusted bounds and bounds under noncompliance but only among the compliers under the monotonicity assumption. \citet{huang2016inequality} consider randomized trials and derive bounds for the probability of benefit with and without covariate adjustment, allowing for any restriction on the joint distribution of the potential outcomes, e.g., no harm or no harm at certain levels of the outcome. \citet{fay2018causal} discuss the Mann-Whitney statistic and the proportion who benefit adjusted for ties, the individual-level version of the Mann-Whitney, and contrasts these two estimands via Hand's paradox \citep{hand1992comparing}. They report valid bounds constructed from the previously derived bounds of \citet{fan2010sharp, huang2016inequality} in the continuous case and \citet{lu2018treatment} in the ordinal case. These results apply in settings where the marginal distributions of the potential outcomes are known or can be estimated, i.e., in randomized trials. \citet{lu2020sharp} derive sharp bounds for the relative treatment effect, i.e., the proportion who benefit minus the proportion harmed, which has a one-to-one relation to the estimand discussed in \citet{fay2018causal}, where the bounds expressions depend on marginal distributions of potential outcomes. \citet{lu2020sharp} point out that although \citet{fay2018causal} claim that all their bounds are tight, they are, in fact, only valid. 

Although there has been a long literature on bounds for the probability of benefit and variations on it, bounds in the noncompliance setting have to our knowledge only been derived for the probability of benefit and of no harm for compliers under the no defiers (monotonicity) assumption \citep{lu2018treatment}. Finally, bounds have not been, to our knowledge, derived under the observational setting with unmeasured confounders, which would be the worst-case scenario in an observational study. All of the bounds we derive are in terms of probabilities of observable quantities given the setting, i.e., not involving potential outcome probabilities. 

We derive bounds for all three estimands considered in the literature, the probability of no harm, the probability of benefit, and the relative treatment effect in at least two novel settings for each estimand for up to an eight-level ordinal outcome measure. Additionally, in the observational setting with unmeasured confounders, we derive a general set of bounds for each estimand allowing for an ordinal outcome of arbitrary, but finite, levels. In the noncompliance setting, we consider monotonicity assumptions and compare, when possible, to previously published bounds. In doing so, we find, as did \citet{lu2020sharp}, that the bounds reported in \citet{fay2018causal} are not, as presented, sharp. We discuss the reason for this lack of sharpness due to how the bounds were constructed from the addition of two sets of sharp bounds. We provide simulation comparisons of the bounds' widths under different probability law scenarios and different Directed Acycle Graph (DAG) settings. Finally, we provide a data example using the data from the peanut allergy trial \citep{du2015randomized}, which had imperfect compliance and drop-out. We consider only the imperfect compliance complication, leaving the potentially informative drop-out for future research.

\section{Preliminaries} 
Let $X$ denote the intervention or exposure of interest, which will be binary taking values 0, 1, and $Y$ the ordinal outcome of interest that takes one of $K$ levels $0, \ldots, K-1$. Let $Z$ denote a binary instrumental variable, and $U$ be a set of unmeasured variables. Let $Y_i(x)$ be the potential outcome of Y had subject $i$ had treatment or exposure level $X=x$. We are interested in the three settings that are depicted in Figures \ref{DAGS} below. Here we follow \citep{pearl2009causality} such that the DAG in Figure \ref{c} encodes the following nonparametric structural equation models (NPSEM):
\begin{align*}
z = g_Z(\epsilon_Z), \qquad x =g_X(u, z, \epsilon_X), \qquad y = g_Y(u, x, \epsilon_Y)
\end{align*}
for some functions $g_Z, g_X, g_Y$. The set of unmeasured variables $U$ and the set of $\epsilon$s represent errors due to omitted factors. Given the values of the errors and the values of a variable's parents in the graph, the value of the variable is determined by its NPSEM. In Figure \ref{b}, we omit the instrumental variable $Z$ so that instead $x = g_X(u, \epsilon_X)$ while the structural equation for $Y$ remains the same. In Figure \ref{a}, we omit the unmeasured confounder $U$ so that $x = g_X(\epsilon_X)$ and $y = g_Y(x, \epsilon_Y)$ corresponding to a randomized trial setting. 

In Figure \ref{c} we also consider the additional assumption that $g_X(u, z = 0, \epsilon_X) \leq g_X(u, z = 1, \epsilon_X)$, which we refer to as the no defiers assumption. 

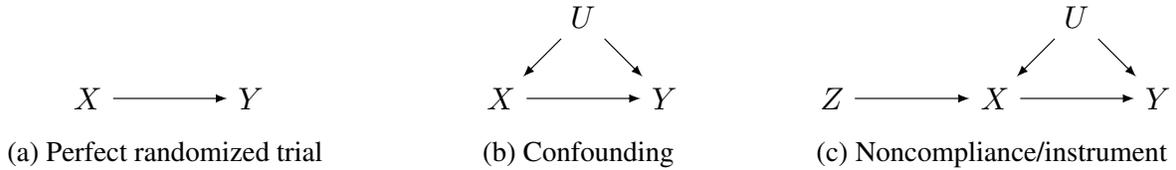
\begin{figure*}[ht]
\captionsetup[sub]{width=.9\linewidth}
\centering
\resizebox{\linewidth}{!}{
\begin{subfigure}[t]{0.3\textwidth}
\centering
\begin{tikzpicture}
\node (v) at (0,0) {$X$};
\node (i) at (2,0) {$Y$};

\draw[-latex] (v) -- (i);

\end{tikzpicture}
\caption{Perfect randomized trial \label{a}}
\end{subfigure}

\begin{subfigure}[t]{0.3\textwidth}
\centering
\begin{tikzpicture}
\node (v) at (0,0) {$X$};
\node (i) at (2,0) {$Y$};
\node (uh) at (1,1) {$U$};
\draw[-latex] (uh) -- (i);

\draw[-latex] (uh) -- (v);
\draw[-latex] (v) -- (i);

\end{tikzpicture}
\caption{Confounding \label{b}}
\end{subfigure}

\begin{subfigure}[t]{0.3\textwidth}
\centering
\begin{tikzpicture}
\node (E) at (-2,0) {$Z$};
\node (uh) at (1,1) {$U$};
\node (v) at (0,0) {$X$};
\node (i) at (2,0) {$Y$};
\draw[-latex] (E) -- (v);


\draw[-latex] (v) -- (i);

\draw[-latex] (uh) -- (i);

\draw[-latex] (uh) -- (v);


\end{tikzpicture}
\caption{Noncompliance/instrument \label{c}}
\end{subfigure}
}
\caption{Causal diagrams for three possible study designs. \label{DAGS}}
\end{figure*}

Under each setting we are interested in bounds for three estimands: the probability of no harm, denoted $\psi$, the probability of benefit, denoted $\theta$, and the relative treatment effect, denoted $\phi$. These are defined as:
\begin{align*}
    \psi &= \P\{Y_i(1) \geq Y_i(0)\},\\
    \theta &=\P\{Y_i(1) > Y_i(0)\},\\
    \phi &= \P\{Y_i(1) > Y_i(0)\} - \P\{Y_i(1) < Y_i(0)\}.
\end{align*}
The estimand $\phi$ is a transformation of the estimand considered in \citet{fay2018causal}, which is the translation of the Mann-Whitney statistic to the individual level causal effect, $\P\{Y_i(1)>Y_i(0)\} + 2^{-1}\P\{Y_i(1)=Y_i(0)\}$, which was shown to be equal to $2^{-1}+2^{-1}\phi$. 

\subsection{Previous bounds}

\cite{lu2018treatment} derive sharp bounds on $\psi$ and $\theta$ in terms of the marginal potential outcome probabilities $\P\{Y(1) = k\}$ and $\P\{Y(0) = k\}$, which can be estimated in randomized studies with perfect compliance, or if there are no unmeasured confounders. The bounds are
\[
\max_{0 \leq j \leq K - 1}\left[\P\{Y(0) = j\} + \P\{Y(1) \geq j\} - P\{Y(0) \geq j\}\right] \leq \psi \leq 
\]
\[
1 + \min_{0 \leq j \leq K - 1}\left[\P\{Y(1) \geq j\} - \P\{Y(0) \geq j\}\right] 
\]
and 
\[
\max_{0 \leq j \leq K - 1}\left[\P\{Y(1) \geq j\} - \P\{Y(0) \geq j\}\right] \leq \theta \leq 
\]
\[
1 + \min_{0 \leq j \leq K - 1}\left[\P\{Y(1) \geq j\} - \P\{Y(0) \geq j\} - \P\{Y(0) = j\}\right].
\]
\citet{fay2018causal} derive valid bounds on $2^{-1} + 2^{-1} \phi$ using the results of \citet{lu2018treatment}, however, their procedure does not yield sharp bounds. \citet{lu2020sharp} derived sharp bounds on $\phi$ again in terms of the marginal potential outcome probabilities. The upper bound on $\phi$ is 
\begin{align*}
\min_{1 \leq j \leq K - 1} \min_{1 \leq m \leq K - j} \left[\sum_{k=j}^{K-1} \P\{Y(1) = k\} + \sum_{k=j + m}^{K-1} \P\{Y(1) = k\} + \right.\\ 
\left.\sum_{l=0}^{j-2} \P\{Y(0) = l\} - \sum_{l=j + m - 1}^{K-1} \P\{Y(0) = l\} \right]
\end{align*}
and the lower bound on $\phi$ is 
\begin{align*}
\max_{1 \leq j \leq K - 1} \max_{1 \leq m \leq K - j} \left[\sum_{k=j + m - 1}^{K-1} \P\{Y(1) = k\} - \sum_{l=j}^{K-1} \P\{Y(0) = l\} - \right. \\
\left.\sum_{l=j + m}^{K-1} \P\{Y(0) = l\} - \sum_{k=0}^{j-2} \P\{Y(1) = k\} \right].
\end{align*}

\section{Novel bounds}
We derive bounds for $\psi$, $\theta$ and $\phi$ in each of the settings shown in Figure \ref{DAGS} using the method of \citet{sachs2022general} and as implemented in the R package \texttt{causaloptim}. In the setting of Figure \ref{a}, the response functional for $Y$ has a discrete codomain with $K^2$ levels since it only depends on $x$. In the other settings, since $X$, $Y$, and $Z$ are all assumed categorical, there exists a canonical partitioning of the unmeasured confounder $U$ into finite states, as described by \citet{sachs2022general}. In this partitioning, the response functional for $Y$ still has categorical codomain with $K^2$ levels and the response functional for $X$ has categorical codomain with 2 levels in the case of Figure \ref{b} and 4 levels in the case of Figure \ref{c}. There are thus a finite set of probabilities associated with the response function variable distribution in all three cases. In particular, there are $2 K^2$ for $Y \vert X$ in Figure \ref{a}, $2 K^2$ for $Y, X$ in Figure \ref{b}, and $2^2 K^2$ for $Y, X \vert Z$ in Figure \ref{c}. The notation used in the bounds is as follows, $p_{xy\cdot z} = \P(X = x, Y = y | Z = z)$ and $p_{xy} = \P(X = x, Y = y)$.

\subsection{Bounds for $\psi$}

As general bounds for $\psi$ under the perfect randomized trials setting have been previously presented in \citet{lu2020sharp}, we consider only the settings in the DAGS \ref{b} and \ref{c} in Figure \ref{DAGS} for a various number of levels for the outcome $Y$.

\begin{theorem}
In the setting of DAG \ref{b} in Figure \ref{DAGS} the tight and valid bounds for $\psi=\P\{Y_i(1) \geq Y_i(0)\}$ for a discrete $Y$ with arbitrary levels $K$ is given by: 
$$p_{00} + p_{1(K-1)} \leq\psi\leq 1.$$

\noindent Proof of Theorem 1 is given in the supplementary materials. 
\end{theorem}

\begin{result}
In the setting of DAG \ref{c} in Figure \ref{DAGS} the tight and valid bounds for $\psi=\P\{Y_i(1) \geq Y_i(0)\}$, for a discrete $Y$ with three levels $\{0,1,2\}$ is given by: 
\[ 
 \psi \geq \mbox{max} \left. \begin{cases}   1 - p_{10\cdot 0} - p_{01\cdot 0} - p_{11\cdot 0} - p_{02\cdot 0},\\ 
   1 - p_{10\cdot 0} - p_{11\cdot 0} - p_{02\cdot 0} - p_{10\cdot 1} - p_{01\cdot 1},\\ 
   1 - p_{10\cdot 1} - p_{01\cdot 1} - p_{11\cdot 1} - p_{02\cdot 1},\\ 
   1 - p_{11\cdot 0} - p_{02\cdot 0} - p_{10\cdot 1} - p_{01\cdot 1} - p_{02\cdot 1},\\ 
   1 - p_{10\cdot 0} - p_{01\cdot 0} - p_{02\cdot 0} - p_{11\cdot 1} - p_{02\cdot 1},\\ 
   1 - p_{10\cdot 0} - p_{01\cdot 0} - p_{10\cdot 1} - p_{11\cdot 1} - p_{02\cdot 1},\\ 
   1 - p_{00\cdot 0} - p_{10\cdot 0} - p_{01\cdot 0} - 2p_{11\cdot 0} - 2p_{02\cdot 0} + p_{11\cdot 1},\\ 
   1 + p_{11\cdot 0} - p_{00\cdot 1} - p_{10\cdot 1} - p_{01\cdot 1} - 2p_{11\cdot 1} - 2p_{02\cdot 1},\\ 
   p_{00\cdot 0} - p_{10\cdot 0} - p_{01\cdot 0} + p_{01\cdot 1},\\ 
   p_{01\cdot 0} + p_{00\cdot 1} - p_{10\cdot 1} - p_{01\cdot 1} \end{cases} \right\} 
 \] \[ 
 \psi \leq \mbox{min} \left. \begin{cases}   1 + p_{00\cdot 0} - p_{10\cdot 0} + p_{01\cdot 0} + p_{11\cdot 0} + p_{00\cdot 1} + p_{10\cdot 1} - p_{11\cdot 1},\\ 
   2 - p_{00\cdot 0} - p_{10\cdot 0} - p_{01\cdot 1} - p_{02\cdot 1},\\ 
   1 - p_{10\cdot 0} + p_{00\cdot 1} + p_{10\cdot 1},\\ 
   2 - p_{10\cdot 0} - p_{02\cdot 0} - p_{01\cdot 1} - p_{11\cdot 1},\\ 
   3 - p_{00\cdot 0} - p_{10\cdot 0} - 2p_{02\cdot 0} - p_{00\cdot 1} - p_{10\cdot 1} - 2p_{01\cdot 1} - p_{11\cdot 1},\\ 
   2 - p_{00\cdot 0} - p_{10\cdot 0} - p_{02\cdot 0} - p_{01\cdot 1},\\ 
   1 + p_{00\cdot 0} + p_{10\cdot 0} - p_{11\cdot 0} + p_{00\cdot 1} - p_{10\cdot 1} + p_{01\cdot 1} + p_{11\cdot 1},\\ 
   1 + p_{00\cdot 0} + p_{10\cdot 0} - p_{10\cdot 1},\\ 
   2 - p_{01\cdot 0} - p_{02\cdot 0} - p_{00\cdot 1} - p_{10\cdot 1},\\ 
   2 - p_{01\cdot 0} - p_{11\cdot 0} - p_{10\cdot 1} - p_{02\cdot 1},\\ 
   2 - p_{01\cdot 0} - p_{00\cdot 1} - p_{10\cdot 1} - p_{02\cdot 1},\\ 
   2 - p_{02\cdot 0} - p_{00\cdot 1} - p_{10\cdot 1} - p_{01\cdot 1} - p_{11\cdot 1},\\ 
   1 + p_{00\cdot 0} + p_{10\cdot 0} + p_{01\cdot 0} + p_{11\cdot 0} - p_{10\cdot 1} - p_{11\cdot 1},\\ 
   1 + p_{00\cdot 0} + p_{01\cdot 0} + p_{11\cdot 0} - p_{11\cdot 1},\\ 
   1 - p_{11\cdot 0} + p_{00\cdot 1} + p_{01\cdot 1} + p_{11\cdot 1},\\ 
   2 - p_{00\cdot 0} - p_{10\cdot 0} - p_{01\cdot 0} - p_{11\cdot 0} - p_{02\cdot 1},\\ 
   1 - p_{10\cdot 0} - p_{11\cdot 0} + p_{00\cdot 1} + p_{10\cdot 1} + p_{01\cdot 1} + p_{11\cdot 1},\\ 
   3 - p_{00\cdot 0} - p_{10\cdot 0} - 2p_{01\cdot 0} - p_{11\cdot 0} - p_{00\cdot 1} - p_{10\cdot 1} - 2p_{02\cdot 1}\\
1\end{cases} \right\} 
 \]
\end{result}

\begin{result}
In the setting of DAG \ref{c} in Figure \ref{DAGS} with the additional assumption of no defiers, the tight and valid bounds for $\psi=\P\{Y_i(1) \geq Y_i(0)\}$, for a discrete $Y$ with three levels $\{0,1,2\}$ is given by: 

\[ 
\psi \geq \mbox{max} \left. \begin{cases}   1 - p_{10\cdot 1} - p_{01\cdot 1} - p_{11\cdot 1} - p_{02\cdot 1},\\ 
   1 - p_{11\cdot 0} - p_{02\cdot 0} - p_{10\cdot 1} - p_{01\cdot 1},\\ 
   1 - p_{10\cdot 0} - p_{01\cdot 0} - p_{11\cdot 0} - p_{02\cdot 0} \end{cases} \right\} 
 \] \[ 
\psi \leq \mbox{min} \left. \begin{cases}   1 + p_{00\cdot 0} + p_{10\cdot 0} - p_{00\cdot 1} - p_{10\cdot 1},\\ 
   1 + p_{00\cdot 0} + p_{10\cdot 0} + p_{01\cdot 0} + p_{11\cdot 0} - p_{00\cdot 1} - p_{10\cdot 1} - p_{01\cdot 1} - p_{11\cdot 1},\\
1\end{cases} \right\} 
 \]
\end{result}

We also derived symbolic bounds for $\psi$ in the settings of DAGs \ref{c} in Figure \ref{DAGS} for a discrete $Y$ with up to $K = 7$ levels, without and with the assumption of no defiers. We do not present these expressions because they are long, complex, and are difficult to glean insight from. Instead, we provide R functions in the supplementary materials that can be used to compute the bounds in these settings given the observable probabilities.

\subsection{Bounds for $\theta$}

As general bounds for $\theta$ under the perfect randomized trials setting have been previously presented in \citet{lu2018treatment}, we consider only the settings in the DAGs \ref{b} and \ref{c} in Figure \ref{DAGS}. 

\begin{theorem}
In the setting of DAG \ref{b} in Figure \ref{DAGS} the tight and valid bounds for $\theta=\P\{Y_i(1) > Y_i(0)\}$, for a discrete $Y$ with arbitrary levels $K$ is given by: 
$$0 \leq\theta\leq 1-p_{10} -p_{0(K-1)}.$$

\noindent Proof of Theorem 2 is given in the supplementary materials. 
\end{theorem}

\begin{result}
In the setting of DAG \ref{c} in Figure \ref{DAGS} the tight and valid bounds for $\theta=\P\{Y_i(1) > Y_i(0)\}$, for a discrete $Y$ with three levels $\{0,1,2\}$ is given by: 

\[ 
 \theta \geq \mbox{max} \left. \begin{cases}   -p_{00\cdot 0} - p_{10\cdot 0} - 2p_{01\cdot 0} - 2p_{11\cdot 0} - 2p_{02\cdot 0} + p_{00\cdot 1} + p_{10\cdot 1} + p_{01\cdot 1} + 2p_{11\cdot 1},\\ 
   -p_{00\cdot 0} - p_{10\cdot 0} - p_{01\cdot 0} - p_{11\cdot 0} + p_{00\cdot 1} + p_{01\cdot 1},\\ 
   -p_{00\cdot 0} - p_{10\cdot 0} - p_{01\cdot 0} - p_{11\cdot 0} - p_{02\cdot 0} + p_{00\cdot 1} + p_{10\cdot 1} + p_{01\cdot 1} + p_{11\cdot 1},\\ 
   -p_{10\cdot 0} - p_{01\cdot 0} - p_{11\cdot 0} - p_{02\cdot 0} + p_{01\cdot 1} + p_{11\cdot 1},\\ 
   p_{00\cdot 0} - p_{10\cdot 0} - p_{01\cdot 0} - p_{11\cdot 0} - p_{00\cdot 1} - p_{10\cdot 1} + p_{01\cdot 1},\\ 
   -p_{10\cdot 0} - p_{01\cdot 0} - p_{11\cdot 0} + p_{01\cdot 1},\\ 
   p_{00\cdot 0} + p_{10\cdot 0} + p_{01\cdot 0} + p_{11\cdot 0} - p_{00\cdot 1} - p_{10\cdot 1} - p_{01\cdot 1} - p_{11\cdot 1} - p_{02\cdot 1},\\ 
   p_{00\cdot 0} + p_{01\cdot 0} - p_{00\cdot 1} - p_{10\cdot 1} - p_{01\cdot 1} - p_{11\cdot 1},\\ 
   p_{01\cdot 0} - p_{10\cdot 1} - p_{01\cdot 1} - p_{11\cdot 1},\\ 
   p_{00\cdot 0} - p_{00\cdot 1} - p_{10\cdot 1},\\ 
   p_{00\cdot 0} + p_{10\cdot 0} - p_{00\cdot 1} - p_{10\cdot 1} - p_{01\cdot 1} - p_{02\cdot 1},\\ 
   0,\\ 
   -p_{01\cdot 0} - p_{11\cdot 0} - p_{02\cdot 0} + p_{11\cdot 1},\\ 
   p_{11\cdot 0} - p_{01\cdot 1} - p_{11\cdot 1} - p_{02\cdot 1},\\ 
   p_{00\cdot 0} + p_{10\cdot 0} + p_{01\cdot 0} + 2p_{11\cdot 0} - p_{00\cdot 1} - p_{10\cdot 1} - 2p_{01\cdot 1} - 2p_{11\cdot 1} - 2p_{02\cdot 1},\\ 
   p_{01\cdot 0} + p_{11\cdot 0} - p_{10\cdot 1} - p_{01\cdot 1} - p_{11\cdot 1} - p_{02\cdot 1},\\ 
   -p_{00\cdot 0} - p_{10\cdot 0} + p_{01\cdot 0} + p_{00\cdot 1} - p_{10\cdot 1} - p_{01\cdot 1} - p_{11\cdot 1},\\ 
   -p_{00\cdot 0} - p_{10\cdot 0} + p_{00\cdot 1},\\ 
   -p_{00\cdot 0} - p_{10\cdot 0} - p_{01\cdot 0} - p_{02\cdot 0} + p_{00\cdot 1} + p_{10\cdot 1} \end{cases} \right\} 
 \] \[ 
 \theta \leq \mbox{min} \left. \begin{cases}   1 - p_{10\cdot 0} - p_{02\cdot 0},\\ 
   2 - p_{10\cdot 0} - p_{01\cdot 0} - p_{02\cdot 0} - p_{00\cdot 1} - p_{10\cdot 1} - p_{11\cdot 1} - p_{02\cdot 1},\\ 
   1 + p_{00\cdot 0} - p_{10\cdot 0} + p_{11\cdot 0} - p_{11\cdot 1},\\ 
   2 - p_{00\cdot 0} - p_{10\cdot 0} - p_{11\cdot 0} - 2p_{02\cdot 0} - p_{01\cdot 1},\\ 
   2 - p_{01\cdot 0} - p_{00\cdot 1} - p_{10\cdot 1} - p_{11\cdot 1} - 2p_{02\cdot 1},\\ 
   1 + p_{00\cdot 0} + p_{11\cdot 0} - p_{10\cdot 1} - p_{11\cdot 1} - p_{02\cdot 1},\\ 
   1 - p_{10\cdot 0} - p_{11\cdot 0} - p_{02\cdot 0} + p_{00\cdot 1} + p_{11\cdot 1},\\ 
   2 - p_{00\cdot 0} - p_{10\cdot 0} - p_{11\cdot 0} - p_{02\cdot 0} - p_{10\cdot 1} - p_{01\cdot 1} - p_{02\cdot 1},\\ 
   1 - p_{10\cdot 1} - p_{02\cdot 1},\\ 
   1 - p_{11\cdot 0} + p_{00\cdot 1} - p_{10\cdot 1} + p_{11\cdot 1} \end{cases} \right\} 
 \]
\end{result}

\begin{result}
In the setting of DAG \ref{c} in Figure \ref{DAGS} with the additional assumption of no defiers, the tight and valid bounds for $\theta=\P\{Y_i(1) > Y_i(0)\}$, for a discrete $Y$ with three levels $\{0,1,2\}$ is given by: 

\[ 
\theta \geq \mbox{max} \left. \begin{cases}   0,\\ 
   p_{00\cdot 0} + p_{10\cdot 0} - p_{00\cdot 1} - p_{10\cdot 1},\\ 
   p_{00\cdot 0} + p_{10\cdot 0} + p_{01\cdot 0} + p_{11\cdot 0} - p_{00\cdot 1} - p_{10\cdot 1} - p_{01\cdot 1} - p_{11\cdot 1} \end{cases} \right\} 
 \] \[ 
\theta \leq \mbox{min} \left. \begin{cases}   1 + p_{00\cdot 0} + p_{11\cdot 0} - p_{00\cdot 1} - p_{10\cdot 1} - p_{11\cdot 1} - p_{02\cdot 1},\\ 
   1 - p_{10\cdot 0} - p_{02\cdot 0},\\ 
   1 - p_{10\cdot 1} - p_{02\cdot 1} \end{cases} \right\} 
 \]
\end{result}

We also derived symbolic bounds for $\theta$ in the settings of DAGs \ref{c} in Figure \ref{DAGS} for a discrete $Y$ with up to $K = 7$ levels without and with the no defiers assumption. We provide R functions in the supplementary materials that can be used to compute the bounds in these settings given the observable probabilities. 

\subsection{Bounds for $\phi$}

As general bounds for $\phi$ under the perfect randomized trials setting has been previously presented in \citet{lu2018treatment}, we consider only the settings in the DAGs \ref{b} and \ref{c} in Figure \ref{DAGS}. 

\begin{theorem}
In the setting of DAG \ref{b} in Figure \ref{DAGS} the tight and valid bounds for $\phi=\P\{Y_i(1) > Y_i(0)\} - \P\{Y_i(1) < Y_i(0)\}$, for a discrete $Y$ with arbitrary levels $K$ is given by: 
$$p_{00} + p_{1(K-1)}-1  \leq\phi\leq 1-p_{10} -p_{0(K-1)}.$$

\noindent Proof of Theorem 3 is given in the supplementary materials. 
\end{theorem}

It is of note, but is not surprising, that when $K=2$, these are the exact bounds for the fully confounded setting in DAG \ref{b} in Figure \ref{DAGS} for the risk difference as first given in \citet{robins1989analysis}. This is because $\phi=\P\{Y_i(1) > Y_i(0)\} - \P\{Y_i(1) < Y_i(0)\} = \P\{Y(1)=1\} - \P\{Y(0)=1\}$ when $K=2$, as stated in \citet{lu2020sharp}. This is easily shown:
\begin{eqnarray*}
\P\{Y_i(1) > Y_i(0)\} &=& \P\{Y_i(1)=1,Y_i(0)=0\}\\
\P\{Y_i(1) < Y_i(0)\} &=& \P\{Y_i(1)=0,Y_i(0)=1\}\\
\P\{Y(1)=1) &=& \P\{Y_i(1)=1,Y_i(0)=0\}+\P\{Y_i(1)=0,Y_i(0)=1\}\\
\P\{Y(0)=1) &=& \P\{Y_i(1)=1,Y_i(0)=1\}+\P\{Y_i(1)=0,Y_i(0)=1\} \mbox{ thus,}\\
\P\{Y(1)=1)-\P(Y(0)=1\} &=& \P\{Y_i(1)=1,Y_i(0)=0\}-
\P\{Y_i(1)=0,Y_i(0)=1\} = \phi.
\end{eqnarray*}

We derive bounds for $\phi$ under the setting of DAG \ref{c} in Figure \ref{DAGS} with $K = 3$ without any additional assumptions. This produces tight and valid bounds with 36 terms in both lower and upper bounds, respectively. For this reason, we do not produce these bounds in the main text. They are given in the supplementary materials.

\begin{result}
In the setting of DAG \ref{c} in Figure \ref{DAGS} with the additional assumption of no defiers, the tight and valid bounds for $\phi=\P\{Y_i(1) > Y_i(0)\} - \P\{Y_i(1) < Y_i(0)\}$, for a discrete $Y$ with three levels $\{0,1,2\}$ is given by: 
\[ 
\phi \geq \mbox{max} \left. \begin{cases}   p_{00\cdot 0} - p_{11\cdot 0} - p_{02\cdot 0} - p_{00\cdot 1} - p_{10\cdot 1} - p_{01\cdot 1},\\ 
   p_{00\cdot 0} - p_{00\cdot 1} - p_{10\cdot 1} - p_{01\cdot 1} - p_{11\cdot 1} - p_{02\cdot 1},\\ 
   p_{00\cdot 0} + p_{10\cdot 0} + p_{01\cdot 0} - p_{00\cdot 1} - 2p_{10\cdot 1} - 2p_{01\cdot 1} - p_{11\cdot 1} - p_{02\cdot 1} \end{cases} \right\} 
 \] \[ 
\phi \leq \mbox{min} \left. \begin{cases}   1 - p_{02\cdot 0} - p_{10\cdot 1},\\ 
   1 + p_{00\cdot 0} + p_{11\cdot 0} - p_{02\cdot 0} - p_{00\cdot 1} - p_{10\cdot 1} - p_{11\cdot 1},\\ 
   1 + p_{00\cdot 0} + p_{10\cdot 0} + p_{11\cdot 0} - p_{00\cdot 1} - 2p_{10\cdot 1} - p_{11\cdot 1} - p_{02\cdot 1} \end{cases} \right\} 
 \]
\end{result}

We also derive symbolic bounds for $\phi$ in the settings of DAGs \ref{c} in Figure \ref{DAGS} for a discrete $Y$ with up to $K = 7$ levels without and with the no defiers assumption. We provide R functions in the supplementary materials to compute the bounds in these settings given the observable probabilities. 

\section{Simulations}

We conduct a simulation study to assess and compare the width of the bounds in the different study designs and with different numbers of levels of the outcome over a range of simulated true probability laws. We also compare ours to the bounds reported in \citet{fay2018causal}, which were derived based on a combination of the bounds in \citet{lu2018treatment}, and to the bounds reported in \citet{lu2020sharp}, in the randomized trial setting with perfect compliance where marginal probabilities on potential outcomes can be directly estimated. These simulations use true probabilities, which are generated under mechanisms that correspond to the causal models in Figure \ref{DAGS}. \\

\textbf{To generate distributions under the perfect randomized trial} setting we randomly sample a probability $p_x \sim \mbox{Uniform}(0.2, 0.8)$, and assume $X_i \sim \mbox{Bernoulli}(p_z)$. We randomly sample a treatment effect $\theta \sim \mbox{Normal}(0, \mbox{sd} = 2)$ and assume a latent continuous variable $\tilde{Y}_i \vert X_i \sim \mbox{Normal}(X_i \theta, 1)$. We assume that the outcome $Y_i$ is a categorical version of $\tilde{Y}_i$ obtained by binning into $K$ bins each with probability mass $1/K$. The true probabilities of the form $\P(Y_i = j \vert X_i = x)$, for $x \in \{0, 1\}, j \in \{0, \ldots, K - 1\}$, are finally obtained from the Normal distribution function. \\

\textbf{To generate distributions under the confounded setting} we assume $U_i \sim \mbox{Normal}(0,1)$, and then randomly sample $a_1, a_2$ from $\mbox{Normal}(0,1)$. We assume $X_i \vert U_i \sim \mbox{Bernoulli}(p_i)$, where $p_i = \Phi(a_1 + a_2 U_i)$, where $\Phi(\cdot)$ is the standard normal cumulative distribution function. Then we randomly sample ${b_1, b_2}$ from $\mbox{Normal}(0, 2)$ and assume $\tilde{Y}_i \sim \mbox{Normal}(X_i \theta + U_i b_1 + U_i X_i b_2, 1).$ We assume that the outcome $Y_i$ is a categorical version of $\tilde{Y}_i$ obtained by binning into $K$ bins each with probability mass $1/K$. The true probabilities of the form $\P(Y_i = j, X_i = x)$, for $x \in \{0, 1\}, j \in \{0, \ldots, K - 1\}$, are obtained by calculating the joint distribution of $Y_i, X_i, U_i$ and then marginalizing out $U_i$. \\

\textbf{To generate distributions under the noncompliance setting}, we need the unmeasured common cause of $X_i$ and $Y_i$ that is also related to how $X_i$ is determined from $Z_i$. In particular, we randomly sample a $p_z \sim \mbox{Uniform}(0.2, 0.8)$ and assume
\begin{eqnarray*}
Z_i &\sim & \mbox{Bernoulli}(p_z), \\
U_i &\sim & \mbox{Normal}(0, 1). 
\end{eqnarray*}
We randomly sample ${b_1, b_2}$ from $\mbox{Normal}(0, 2)$, and $c_1, c_2, c_3$ from Uniform$(0,1)$. Then let $c_{(1)}, c_{(2)}, c_{(3)}$ be the order statistics ($c_1, c_2, c_3$ sorted in increasing order). In order to specify the distribution of $X_i$, let $\bar{U}_i$ follow the distribution obtained by categorizing $U_i$ into four bins with cutpoints at the standard normal percentiles $\Phi^{-1}\{c_{(1)}\}, \Phi^{-1}\{c_{(2)}\}, \Phi^{-1}\{c_{(3)}\}$, where $\Phi^{-1}\{\cdot\}$ is the quantile function of the standard normal. The variable $X_i$ is assumed to follow the following response patterns (never takers, always takers, compliers, and defiers): 
\[
X_i = \begin{cases}
0 \mbox{ if } \bar{U}_i \in (-\infty, \Phi^{-1}\{c_{(1)}\}] \\
1 \mbox{ if } \bar{U}_i \in (\Phi^{-1}\{c_{(1)}\}, \Phi^{-1}\{c_{(2)}\}] \\
Z_i \mbox{ if } \bar{U}_i \in (\Phi^{-1}\{c_{(2)}\}, \Phi^{-1}\{c_{(3)}\}] \\
1 - Z_i \mbox{ if } \bar{U}_i \in (\Phi^{-1}\{c_{(3)}\}, \infty] 
\end{cases}.
\]
The continuous latent variable is assumed to be
\[
\tilde{Y}_i \sim \mbox{Normal}(X_i \theta + U_i b_1 + U_i X_i b_2, 1),
\]
and finally the outcome $Y_i$ is assumed to be the categorized version $\tilde{Y}_i$ into $K$ bins each with probability mass $1/K$. True probabilities of the form $\P(Y_i = j, X_i = x \vert Z_i = z)$ are calculated by marginalizing over $U_i$ from the joints $\P(Y_i = j, X_i = x, U_i = u \vert Z_i = z)$, for $j \in \{1, \ldots, K\}, x \in \{0, 1\},$ and $z \in \{0, 1\}$. \\

\textbf{To generate distributions under the noncompliance setting assuming no defiers}, we use the same construction as the noncompliance setting, except $U_i$ is categorized into 3 bins based on $c_{(1)}, c_{(2)}$, and we remove the defiers response pattern: $X_i = 1 - Z_i$. \\

In each of these scenarios, with $K = 3, 4, 5$, and for each of 5000 replicates, we randomly generate parameters $p_z, \theta, b_1, b_2, c_{(1)}, c_{(2)}, c_{(3)}$ anew at each replicate. These parameters determine the distributions described above from which we compute the true probabilities. The true probabilities are used to compute the bounds under that particular law generating mechanism. 

\begin{figure}
    \centering
    \includegraphics[width = .95\textwidth]{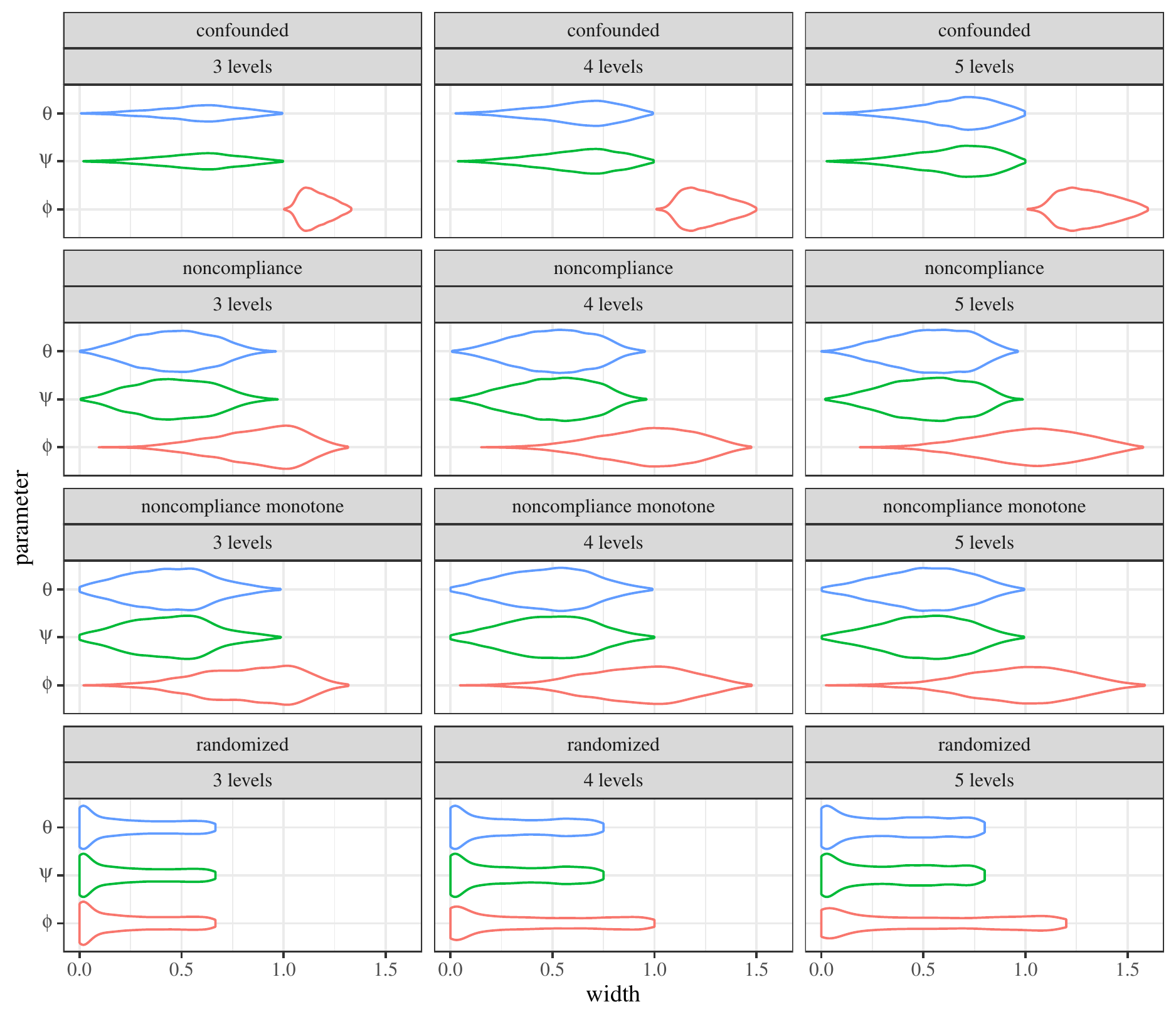}
    \caption{Violin plots showing the distributions of the width of the bounds in different settings. }
    \label{simviolin}
\end{figure}

Figure \ref{simviolin} shows the distributions of the width of the bounds for 5000 simulation replicates in each scenario. The width of the bounds tends to be highest in the confounded setting, as that contains the least amount of information. The noncompliance setting tends to contain more information, but the distribution of the width of bounds is spread out quite a bit. The randomized setting tends to have the shortest width, with a minimum width of nearly 0, and a maximum width of $(K - 2) / K$. As the number of levels of the outcome increases so so both the mean and the spread of the width distributions.

Comparing our bounds to those reported by \citet{lu2020sharp}, we find that they are numerically identical in all cases (over 10,000 replicates). In contrast, our bounds are subintervals, always contained within, the bounds reported by \citep{fay2018causal} in all cases, indicating that the latter are not tight, but are valid.

\section{Data Example}
The results of the trial are presented in \citet{du2015randomized}. The publicly available trial data were downloaded from the Immune Tolerance Network TrialShare website on 2020-06-15 (\url{https://www.itntrialshare.org/}, study identifier: ITN032AD).
The trial had a two armed randomization. However, the trial did not have perfect compliance. In this study, 640 participants between 4 months and 11 months of age were randomized to either consume peanuts or avoid peanuts until the age of 60 months. Those in the consumption arm were instructed to consume at least 6 grams per week. Compliance with the assigned intervention was assessed weekly by using a food frequency questionnaire, and by manual inspection of the infants' cribs for peanut crumbs in a randomly selected subset of participants. The treatment taken was defined as having consumed at least 0.2 grams of peanuts per week. At the end of the study, the outcome, allergy to peanuts, was assessed using an oral food challenge in which infants were given increasing doses of peanuts and their allergic response was assessed. We define the ordinal outcome with 3 levels as the cumulative number of grams of peanuts tolerated at the oral food challenge, categorized into 0 to $<$4 grams (level 0), 4 to $<$8 grams (level 1), and $\geq$8 grams (level 2). 

We compute the bounds in this example for each parameter and under each setting. In the randomized trial setting, we bound the effect of assignment to consume peanuts. In the confounded setting, we bound the effect of actually consuming peanuts, ignoring the randomized assignment. In the noncompliance setting, we bound the effect of actually consuming peanuts using the randomized assignment as the instrumental variable. The results are shown in Table \ref{table:example}.  

\begin{table}[ht]
\centering
\begin{tabular}{p{3cm}llp{3.5cm}}
 & $P\{Y_i(1) \geq Y_i(0)\}$ & $P\{Y_i(1) > Y_i(0)\}$ & $P\{Y_i(1) > Y_i(0)\} - P\{Y_i(1) < Y_i(0)\}$ \\ 
  \hline
assignment effect & (0.68, 1.00) & (0.10, 0.50) & (0.06, 0.21) \\ 
\hline
\hline
  treatment effect ignoring assignment & (0.25, 1.00) & (0.00, 0.84) & (-0.75, 0.84) \\ 
  treatment effect & (0.69, 1.00) & (0.10, 0.49) & (0.05, 0.22) \\ 
  treatment effect assuming no defiers & (0.69, 1.00) & (0.10, 0.49) & (0.05, 0.22) \\ 
  \hline
\end{tabular}
\caption{Lower and upper bounds on treatment effects in the peanut trial under different settings. \label{table:example}}
\end{table}

\section{Discussion} 


We have provided the bounds, the simulation results, and the findings from the real data example in terms of true observed joint probabilities. In practice, probabilities are usually estimated from data. Although estimation of the bounds from data is straightforward using proportions, it would also be desirable to have inference or confidence intervals for the estimated bounds. As noted in \cite{gabriel2022sharp}, this can be done with the nonparametric bootstrap, as long as you are careful about boundary conditions. Improving the method for calculating confidence intervals for bounds in this and many other settings is an area of open research that is being pursued by the authors. Although we have derived bounds under the DAGS in Figure \ref{DAGS} that have no additional measured covariates, all bounds we provide hold and are tight in settings conditional on additional covariates under all DAGS that maintain the causal structure \citep{cai2007non, cai2008bounds}. The efficient incorporation of measured covariates into the bounds is an open area of research for the authors.

\bibliographystyle{plainnat}
\bibliography{ordinal}

\end{document}